\documentclass[runningheads]{llncs}
\usepackage{cite}

\usepackage{graphicx}
\usepackage{xcolor}
\usepackage[T1]{fontenc}

\begin{document}

\title{Neural Feature Selection for Learning to Rank}

\author{Alberto Purpura\inst{1}\thanks{Work done as part of Apple internship.} \and Karolina Buchner\inst{2} \and Gianmaria Silvello\inst{1}\thanks{Work supported by the ExaMode project, as part of the European Union Horizon 2020 program under Grant Agreement no. 825292.} \and \\ Gian Antonio Susto\inst{1}}

\institute{University of Padua, \email{\{purpuraa, silvello, sustogia\}@dei.unipd.it} \and Apple, \email{kbuchner@apple.com}}

\authorrunning{Purpura et al.}


\maketitle              
\begin{abstract}
LEarning TO Rank (LETOR) is a research area in the field
of Information Retrieval (IR) where machine learning models are employed to rank a set of items. In the past few years, neural LETOR approaches have become a competitive alternative to traditional ones like LambdaMART. However, neural architectures performance grew proportionally to their complexity and size. This can be an obstacle for their adoption in large-scale search systems where a model size impacts latency and update time. For this reason, we propose an architecture-agnostic approach based on a neural LETOR model to reduce the size of its input by up to 60\% without affecting the system performance. This approach also allows to reduce a LETOR model complexity and, therefore, its training and inference time up to 50\%.

\keywords{Learning to Rank \and Feature Selection \and Deep Learning}
\end{abstract}
\setcounter{footnote}{0} 
\section{Introduction}
LEarning TO Rank (LETOR) is a research area in the field of Information Retrieval (IR) where machine learning techniques are applied to the task of ranking a set of items \cite{liu2011learning}.
The input to a LETOR system is a set of real-valued vectors representing the items to be ranked -- in decreasing order of relevance -- in return to a certain user query. The output of such systems is usually a set of relevance scores -- one for each item in input -- which estimate the relevance of each item and are used to rank them. In the recent years, the attention on neural approaches for this task has grown proportionally to their performance. Starting from~\cite{ai2018learning}, where the authors propose to employ a recurrent neural layer to model documents list-wise interactions, to~\cite{pobrotyn2020context}, where the now popular self-attention transformer architecture is used. Also, the performance of neural models ~\cite{pobrotyn2020context, zhuang2020feature} recently became competitive with approaches such as LambdaMART~\cite{burges2010ranknet} which is often one of the first choices for LETOR tasks.
However, neural models performance grew at the expense of their complexity and this hampers their application in large-scale search systems. Indeed, in such context, model latency and update time are as important as model performance.
Reducing the input size can help decreasing model architectural complexity, number of parameters, and consequently training and inference time. Also, previous works~\cite{gigli2016fast, geng2007feature, han2018feature} showed that the document representations used for LETOR can sometimes be redundant and often reduced~\cite{gigli2016fast} without impacting the ranking performance. 

Existing feature selection approaches can be organized into three main groups: \textit{filter}, \textit{embedded}, and \textit{wrapper} methods~\cite{gigli2016fast}.\footnote{We purposely omit a comparison with other dimensionality reduction approaches such as PCA since these methods often compute a \emph{combination} of the features to reduce the representation size which is beyond the scope of this paper.} Filter methods, such as the Greedy Search Algorithm (GAS) \cite{geng2007feature}, compute one score for each feature -- independently from the LETOR model that is going to be used afterwards -- and select the top ones according to it. In GAS the authors minimize feature similarity (Kendall Tau) and maximize feature importance. They rank the input items using only one of the features at a time and consider as importance score the MAP or nDCG@k value. Embedded approaches, such as the one presented in \cite{rahangdale2019deep}, incorporate the feature selection process in the model. In \cite{rahangdale2019deep}, the authors propose to apply different types of regularizations -- such as L1 norm regularization 
-- on the weights of a neural LETOR model to reduce redundancy in the hidden representations of the model and improve its performance. Finally, wrapper methods such as the ones presented in \cite{gigli2016fast} and the proposed approach, rely on a LETOR model to estimate feature importance and then perform a selection.

We reimplemented the two best-performing approaches proposed in \cite{gigli2016fast} and consider them as our baselines: eXtended naive Greedy search Algorithm for feature Selection (XGAS) -- which relies on LambdaMART to estimate feature relevance -- and Hierarchical agglomerative Clustering Algorithm for feature Selection (HCAS) employing single likage~\cite{gower1969minimum} -- which relies on Spearman's correlation coefficient between feature pairs as a proxy for feature importance.
To the best of our knowledge, our approach is the first feature selection technique for LETOR specifically targeted to neural models. The main contributions of this paper are the following:
\begin{itemize}
    \item we propose an architecture-agnostic Neural Feature Selection (NFS) approach which uses a neural LETOR model to estimate feature importance;
    \item we evaluate the quality of our approach on two public LETOR collections;
    \item we confirm the robustness of the extracted feature set evaluating the performance of the proposed neural reranker and of a LambdaMART model using subsets of features of different sizes computed with the proposed approach.
\end{itemize}
Our experimental results show that the document representations used for LETOR can sometimes be redundant and reduced to up to 40\% \cite{gigli2016fast} of the total without impacting the ranking performance.
\section{Proposed Approach}
The proposed Neural Feature Selection (NFS) approach is organized in the following three steps. We first train a neural model for the LETOR task, i.e. to compute a relevance score for each item in the input set to be used to rank it. Second, we use the trained model to extract the most significant features groups considered by the model to rank each item. Finally, we perform feature selection using the previously computed feature information.

\begin{figure}[h!]
    \vspace{-15pt}
    \centering
    \includegraphics[scale=0.4]{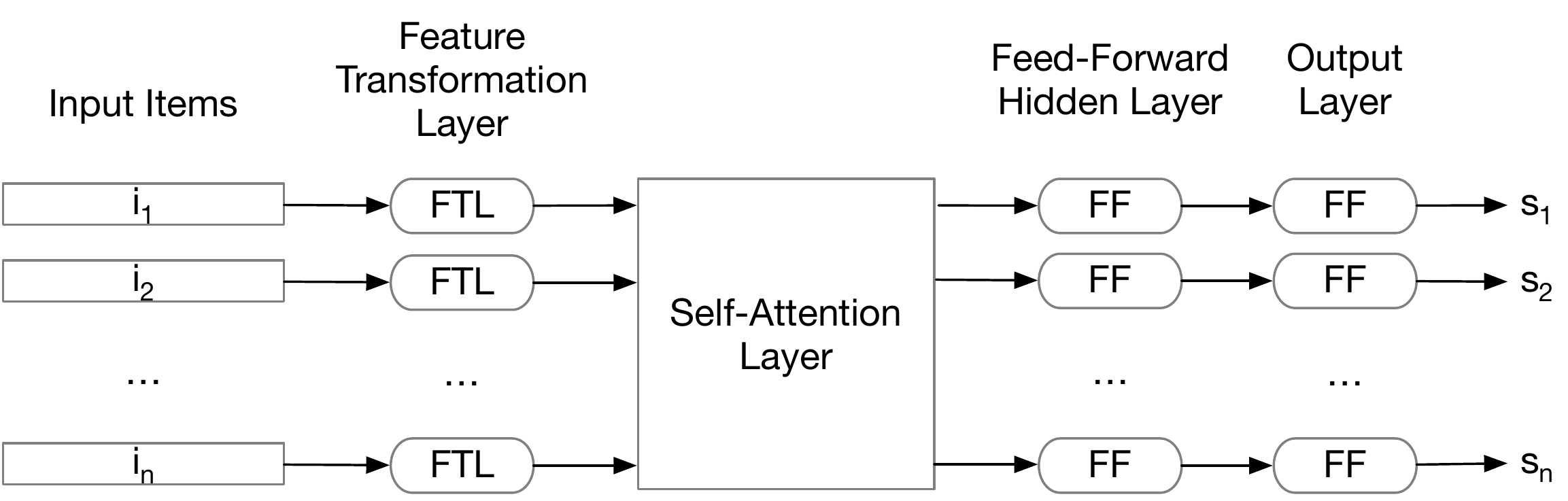}
    \caption{Architecture of the neural architecture employed in our evaluation.}
    \label{fig:arch}
    \vspace{-25pt}
\end{figure}

\paragraph{\textbf{Neural Model Training.}}
The NFS model architecture is composed of $n$ self-attention layers \cite{vaswani2017attention}, followed by two fully-connected layers. We train this model using the ApproxNDCG loss \cite{bruch2019revisiting}. Before feeding the document vectors to the self-attention layer we apply the same feature transformation strategy described in \cite{zhuang2020feature}. In \cite{zhuang2020feature}, the authors apply three different feature transformations to each feature in the input data and then combine them through a weighted sum. The weights for each transformation are learned by the model so that the best feature transformation strategy for each feature could be used each time. The model architecture is depicted in Figure \ref{fig:arch}. Also, we apply batch normalization to the input of each feed-forward layer and dropout on the output of each hidden layer. Note that, since our approach for feature selection is \emph{architecture-agnostic}, we can easily make changes to this neural architecture without impacting the following steps for feature selection.

\paragraph{\textbf{Feature Groups Mining.}}
At this step, we use the model trained in the previous step to select the most important features used to rank each item in our training data. To do so, we compute the saliency map -- a popular approach in the computer vision field to understand model predictions \cite{adebayo2018sanity, shrikumar2016not, simonyan2013deep} -- i.e. the gradient w.r.t. the each input item feature, corresponding to each item in the training dataset. We then apply min-max normalization on each saliency map $M_i$
to map the values in each vector to the same range $[0, 1]$. Afterwards, we select from each saliency map the groups of features $g$ which have a \textit{saliency} score higher than a threshold $t$. The set of feature groups $G$ extracted at this step are the most significant features sets that our neural model learned to rely on to compute the relevance score of each item. These features however might not be the same for any possible input instance and -- as also pointed out in \cite{adebayo2018sanity} -- saliency maps can often be noisy and not always represent the behavior of a neural model. For this reason, we propose to apply a further selection step to prune less reliable feature groups similarly to what proposed in \cite{tonon2019permutation} where the authors compute the statistical significance of groups of items by comparing their frequency of occurrence in real data to the one in randomly generated datasets. 
We compute $K$ random sets of saliency maps, each of the same cardinality of the experimental dataset employed. For example, if a dataset contains $N$ queries, each with $R$ documents to be ranked, then we will generate $K$ random datasets, each containing $N \times R$ saliency maps. Then, we apply the same feature groups extraction process on the random saliency maps and compute $K$ different sets of feature groups. The saliency maps are computed sampling values from a uniform distribution with support $[0, 1]$. According to this modeling strategy, each feature can be considered as salient in the current random saliency map with probability $1-t$; where $t$ is the threshold we used in the previous step to select salient features.
Once we computed these $K$ sets of random feature groups $\hat{G}_k$ we use their frequency to prune the original ones. In particular, we consider the frequency $f_{g_{i}}$ of group $g_i \in G$ and compare it to its frequency in each of the $K$ random datasets $f_{g_{i, k}}$ -- the frequency $f_{g_{i, k}}$ might also be 0 if the feature group $g_i$ does not appear in the random dataset $k$. If $f_{g_{i}} \leq f_{g_{i, k}}$ in more than 2\%~\footnote{This value was set empirically to yield a reasonable number of feature groups for the following feature extraction step.} of the randomly generated feature groups $\hat{G_k}$, we discard feature group $g_i$, considering it as noise.

\paragraph{\textbf{Feature Selection.}}
In this final step, we rely on the feature groups extracted in the previous step and their frequency in the saliency maps to compute a feature similarity matrix. We then use this similarity matrix to perform feature selection.
Each feature pair similarity value is computed counting the times the two features appear in the same feature group and normalizing that score by the total number of groups where that feature appears.
Finally, we rely on this similarity matrix to perform hierarchical clustering as done in \cite{gigli2016fast}. We consider the number of clusters as the stopping criterion for the single linkage hierarchical clustering algorithm. The final set of features to keep is computed selecting the most frequently occurring feature in the previously computed feature groups, from each feature cluster.
\section{Experimental Setup}
We evaluate our approach on the first fold of the MSLR-WEB30K \cite{mslr30k} and on the whole OHSUMED~\cite{qin2010letor} dataset where the items to rank are represented by 136 and 45 features, respectively.~\footnote{\url{https://www.microsoft.com/en-us/research/project/letor-learning-rank-information-retrieval}}
We use the LambdaMART implementation available in the LightGBM \footnote{\url{https://github.com/microsoft/LightGBM}} library \cite{ke2017lightgbm} and train and test the proposed neural model considering only the top 128 results returned by LambdaMART. We tuned the LightGBM model parameters on the validation sets of both datasets, optimizing the ndcg@3 metric.\footnote{We set the learning rate to 0.05, the number of leaves to 200 and the number of trees to 1000 (500) on the MSLR-WEB30K (OHSUMED) collection.} The proposed neural reranking model is trained for 500 epochs -- 100 epochs on the OHSUMED dataset -- with batch size 128, using Adam optimizer and a learning rate 0.0005. We consider a feature embedding size of 128 in the feature transformation layer on the MSLR-WEB30K dataset -- while we removed it for the experiments on the OHSUMED collection due to its much smaller size and number of features which limited the benefits of it -- 4 self-attention heads on the MSLR-WEB30K and 1 on the OHSUMED dataset and a hidden size of 128 for the hidden feed-forward layer. Since each attention head has an output size equal to the total number of features divided by the number of attention heads, to compute the results reported in Table \ref{tab:lm30k}, we reduce the number of attention heads to 1 when using 5\% and 10\% of all the available features (6 and 13 features respectively), we use 4 attention heads when considering 30\% (27 features), and 3 when using 40\% (54 features). The batch normalization momentum we use is 0.4 and the dropout probability is $p=0.5$. In the feature groups mining step, we generate 5000 random datasets and the threshold $t$ to extract the feature groups is empirically set to 0.95. For the evaluation of the approach we consider the nDCG@3 measure, similar results are obtained with nDCG at different cutoffs.
\section{Experimental Results}
In Table \ref{tab:lm30k}, we report the results of our experiments on the MSLR-WEB30K dataset. We trained both a LambdaMART model and the proposed neural reranking one on different subsets of features of increasing size.
From these experiments, we observe that the proposed Neural Feature Selection (NFS) approach always outperforms all the other baselines when the selected features are used to train a LambdaMART model, and in most of the cases when used with the proposed neural model.
\begin{table}[h!]
\scriptsize
\vspace{-15pt}
\centering
\begin{tabular}{p{1.5cm}|p{1.5cm}|p{1.5cm}|p{1.5cm}||p{1.5cm}|p{1.5cm}|p{1.5cm}}
\multicolumn{4}{c||}{LambdaMART} & \multicolumn{3}{c}{Neural Reranker}\\\hline
Features Perc. & XGAS & HCAS (single) & NFS & XGAS & HCAS (single) & NFS \\ \hline\hline
5\%   & 0.3580 & 0.3589 & \textbf{0.3753} & \textbf{0.3768} & 0.3595 & 0.3749 \\
10\%  & 0.3701 & 0.4044 & \textbf{0.4195} & 0.3826 & 0.3923 & \textbf{0.4117} \\
20\%  & 0.3781 & \textbf{0.4672} & \textbf{0.4672} & 0.3831 & \textbf{0.4444} & 0.4434 \\
30\%  & 0.4169 & 0.4655 & \textbf{0.4713} & 0.4085 & \textbf{0.4478} & 0.4236 \\
40\%  & 0.4387 & 0.4709 & \textbf{0.4730} & 0.3943 & 0.4516 & \textbf{0.4559} \\ \hline
100\% &  0.4731 & 0.4731 &  0.4731 & 0.4526 & 0.4526 & 0.4526 \\ \hline
\end{tabular}
\caption{Evaluation of the proposed Neural Feature Selection (NFS) approach on the MSLR-WEB30K dataset. We report the ndcg@3 values obtained by LambdaMART and the proposed Neural Reranking model employing different subsets of features.}
\label{tab:lm30k}
\vspace{-20pt}
\end{table}
The evaluation results on the OHSUMED dataset reported in Table \ref{tab:lmoh} are computed as the previous case. Here, we consider 60\%, 70\%, 80\%, and 90\% of the total features in the collection since the total number of feature is much smaller than in the previous dataset. In our evaluation, NFS outperforms HCAS in the majority of the cases, even though the latter approach is slightly more competitive than before.

The main advantage of using a subset of features to represent the inputs to a neural model is that we can reduce the model complexity. We observe this effect mainly when our data is represented by a large number of features as in the MSLR-WEB30K collection. For example, when using 40\% of the features of the dataset, the number of attention heads in our model was reduced from 4 to 3 and, since we were considering only 54 out of 136 features, the number of parameters of the self-attention heads -- the first layer of our model -- was also reduced. As a consequence, training time was halved and inference time also decreased.
\begin{table}[h!]
\scriptsize
\vspace{-15pt}
 \centering
\begin{tabular}{p{1.5cm}|p{1.5cm}|p{1.5cm}|p{1.5cm}||p{1.5cm}|p{1.5cm}|p{1.5cm}}
\multicolumn{4}{c||}{LambdaMART} & \multicolumn{3}{c}{Neural Reranker}\\\hline
Features Perc. & XGAS & HCAS (single) & NFS & XGAS & HCAS (single) & NFS \\ \hline\hline
 60\%  & 0.3669 & 0.3781 & \textbf{0.3950} & 0.4210 & \textbf{0.4275} & 0.4242 \\
 70\%  & 0.3669 & 0.3781 & \textbf{0.3860} & 0.4243 & 0.4431 & \textbf{0.4437} \\
 80\%  & 0.3669 & 0.3993 & \textbf{0.4007}  & 0.4374 & \textbf{0.4369} & 0.4205 \\
 90\%  & 0.3669 & \textbf{0.4050} & 0.3959 & 0.3669 & 0.4050 & \textbf{0.4221} \\ \hline
 100\% & 0.3968 & 0.3968 & 0.3968 & 0.4973 & 0.4973 & 0.4973 \\ \hline
 \end{tabular}
 \caption{Evaluation of the proposed Neural Feature Selection (NFS) approach on the OHSUMED dataset. We report the ndcg@3 values obtained by LambdaMART and the proposed Neural Reranking model employing different subsets of features.}
 \label{tab:lmoh}
\vspace{-30pt}
\end{table}

It is also interesting to observe the differences between the features selected by the proposed NFS approach and other baselines. We focus on the top 3 features selected from the OHSUMED collection by each of the considered feature selection algorithms over the 5 different dataset folds and refer the reader to \cite{qin2010letor} for a more detailed description of each feature. NFS most frequently selected features computed with popular retrieval models such as BM25 or QLM~\cite{manning2008introduction} (features 4, 12 and 28) based on the document abstract or title. On the other hand, HCAS selected simpler features derived from raw frequency counts of the query terms in each document's title and abstract (features 23, 40 and 36). Finally, XGAS selected a mix of features computed with traditional retrieval approaches such as QL, and simpler frequency counts (features 2, 44 and 13). We conclude that the advantage of NFS is likely due to its ability to recognize and select the most sophisticated and useful matching scores thanks to the information learned during training. 
\section{Conclusions}
In the recent years, neural models became a competitive alternative to traditional Learning TO Rank (LETOR) approaches. Their performance however, grew at the expense of their efficiency and complexity. In this paper, we propose an approach for feature selection for Learning TO Rank (LETOR) based on a neural ranker. Our approach is specifically designed to optimize the performance of neural LETOR models without the need to change their architecture. In our experiments, the proposed approach improved the efficiency of a sample neural LETOR model and decreased its training time without impacting its performance. We also validated the robustness of the selected features testing them using a different -- non neural -- model such as LambdaMART.
We performed our evaluation on two popular LETOR datasets -- i.e. MSLR-WEB30K and OHSUMED -- comparing our approach to three state-of-the-art techniques from \cite{gigli2016fast}. The proposed approach outperformed the selected baselines in the majority of the experiments on both datasets.

%
\newpage
\clearpage
\pagebreak

\bibliographystyle{splncs04}
\bibliography{refs.bib}

\end{document}